# Nonproportional response of LaBr$_3$:Ce and LaCl$_3$:Ce scintillators to synchrotron X-ray irradiation

**Ivan V. Khodyuk and Pieter Dorenbos**
Luminescence Materials Research Group, Faculty of Applied Sciences Delft University of Technology, Mekelweg 15, Delft, 2629 JB, the Netherlands

E-mail: i.v.khodyuk@tudelft.nl

**Abstract.** The nonproportional scintillation response of LaBr$_3$ doped with 5% Ce$^{3+}$ and of LaCl$_3$ doped with 10% Ce$^{3+}$ was measured using highly monochromatic synchrotron irradiation. To estimate the photon response, pulse height spectra at many finely spaced energy values between 9 keV and 100 keV were measured. The experiment was carried out at the X-1 beamline at the Hamburger Synhrotronstrahlungslabor (HASYLAB) synchrotron radiation facility in Hamburg, Germany. Special attention was paid to the X-ray fluorescence escape peaks as they provide us with additional information about photon response in the range 1.2 - 14.5 keV for LaBr$_3$:Ce and 2.0 - 11.6 keV for LaCl$_3$:Ce. A rapid variation of the photon response curve is observed near the Lanthanum K- electron binding energy for both scintillators. A dense sampling of data is performed around this energy and that data are used to apply a method, which we call K-dip spectroscopy. This method allows us to derive the electron response curves of LaBr$_3$:Ce and LaCl$_3$:Ce down to energies as low as 0.1 keV.

## 1. Introduction

*1.1. Theoretical background*

Nonproportional response (nPR) of inorganic scintillators to ionizing radiation is one of the key problems that limits the development of new high energy resolution scintillation detectors [1-5]. The energy resolution $R$, defined as the full width ($\Delta E$) of the full absorption peak in the pulse height spectrum, see figure 1, at half the maximum intensity (FWHM) divided by its energy $E$, of a scintillator detector can be written as [2, 6, 7]

$$\left(\frac{\Delta E}{E}\right)^2 = R^2 = R_{nPR}^2 + R_{inh}^2 + R_p^2 + R_M^2 \quad (1)$$

where $R_{nPR}$ is the contribution of the nonproportional response of the scintillator to the energy resolution, $R_{inh}$ is connected with inhomogeneities in the crystal, which can cause local fluctuations in the scintillation light output, $R_p$ the transfer resolution and $R_M$ the contribution of the photomultiplier tube (PMT) and Poisson statistics in the number of detected photons to the resolution [8, 9]

$$R_M = 2.35\sqrt{(1+\mathrm{var}(M))/N_{phe}^{PMT}} \quad (2)$$

where var$(M)$=0.25 is the contribution from the variance in the gain of the Hamamatsu R6231-100 PMT.

For an absolutely homogeneous scintillation crystal with perfect transfer efficiency, $R_{inh}$ and $R_p$ can be set as zero. Then resolution is given by $R_M$, determined by the variance in the PMT gain and in the number of photoelectrons $N_{phe}^{PMT}$ produced in the PTM, plus $R_{nPR}$. It is important to have $R_{nPR}$ as low as possible to get the best energy resolution at a given $N_{phe}^{PMT}$. To reduce $R_{nPR}$ we need to understand the internal physical cause for the nPR of inorganic scintillators [3, 9-11].

*1.2. Photon and electron nPRs*

In principle, scintillation light yield nonproportionality can be characterized as a function of either photon or electron energy. The scintillation response as a function of X-ray and gamma photon energy, hereafter referred to as the photon nonproportional response (photon-nPR), is in general easy to measure and is an indication of scintillator quality [8]. However, the scintillation nonproportional response as function of electron energy, hereafter referred to as electron nonproportional response (electron-nPR), is more fundamental [4]. For a better understanding of the true cause of nPR, measurements of both the photon and the electron response of the scintillator in question are needed. The most dramatic changes in the nPR occur in the 0.1 keV-10 keV energy range, where the ionization density along the track is higher than at energies of say 100 keV to 1 MeV [1, 12]. To study the nonproportional response in the 0.1 keV – 10 keV range we will apply escape peak analysis and K-dip spectroscopy. These techniques were introduced by us earlier [13, 14].

*1.3. Possible experimental techniques*
To measure photon-nPR, a set of radioactive sources [2, 12] or an energy tunable monochromatic X-ray facility [15, 16] can be used. Figure 1 shows a pulse height spectrum measured by $LaBr_3$:Ce at 45 keV monochromatic X-ray irradiation. From this spectrum we can determine the photon-nPR of $LaBr_3$:Ce at 45 keV. Measuring pulse height spectra at many finely spaced energy values between 9 keV and 100 keV we can determine the entire curve. Due to a short attenuation length of X-rays with energy below 9 keV the scintillator surfaces may affect the scintillation output. It is then difficult to measure the genuine photon-nPR below that energy. Extracting additional data of photon-nPR by analyzing X-ray fluorescence escape peaks [17], gives us information about nonproportionality in the low energy range down to 1 keV. We call this type of nonproportionality curve, escape-nPR [13, 14].
For determining the electron-nPR, the Compton Coincidence Technique (CCT) [18] is a powerful measurement technique. Unfortunately, CCT is not very accurate for the measurement of electron-nPR below 3 keV. An alternative technique that we call K-dip spectroscopy allows us to estimate K-electron-nPR down to energies as low as 70 eV.

*1.4. $LaCl_3$:Ce and $LaBr_3$:Ce*
Discovered in 2001 and 2002, $LaCl_3$:Ce [19, 20] and $LaBr_3$:Ce [21] are among the best scintillators available for X-ray and gamma ray detection [22, 23]. With a high light output of 70000 photons/MeV [24] and low energy resolution of 2.9% observed for the 662 keV full absorption peak, $LaBr_3$:Ce is the benchmark for new potentially high performance scintillation materials [25-27]. $LaCl_3$:Ce with light output of 49000 photons/MeV [24] and energy resolution of 3.3% at 662 keV is a very good scintillator as well [28]. The nonproportional scintillation response of $LaBr_3$:Ce and $LaCl_3$:Ce was measured by several groups [7, 15, 28-30] using various methods down to photon energies of 5 keV and electron energy of 3 keV. In this work we extended those measurements to 1 keV for X-rays and 0.1 keV for electrons. Such data is needed to better understand the true origin of nPR. The main aim of this work is to present new data on photon-nPR and electron-nPR of the scintillators $LaBr_3$:Ce and $LaCl_3$:Ce and to present the new methods used to obtain it.

**2. Experimental methods**

*2.1. $LaBr_3$:Ce and $LaCl_3$:Ce samples*
$LaBr_3$:Ce and $LaCl_3$:Ce are hygroscopic and to study their photon-nPR down to X-ray energies of 9 keV, X-ray assemblies were manufactured by the company Saint-Gobain. Since we intended to exploit X-ray escape peaks for our studies, small 10 mm diameter and 2 mm thick crystals were used to increase the probability of X-ray fluorescence escape. 220 μm thick beryllium was used as an entrance window for the X-rays in order to avoid too much absorption at low energies. The crystals are sealed in aluminum housing with 1 mm thick quartz windows, and the 2 mm edge of the crystal is covered with a white reflector to maximize the photon collection at the PMT photocathode.

The number of photoelectrons $N_{phe}^{PMT}$ per MeV of absorbed energy produced in a Hamamatsu R6231-100 PMT by LaBr$_3$:Ce or LaCl$_3$:Ce was determined by comparing the position of the $^{137}$Cs 662 keV photopeak or of the $^{241}$Am 59.5 keV photopeak in recorded pulse height spectra with the mean value of the so-called single photoelectron pulse height spectrum. The procedure has been described in detail by de Haas *et al.* [24].

*2.2. Synchrotron facility*

To measure the pulse height spectra at many finely spaced energy values between 9 keV and 100 keV, experiments were carried out at the X-1 beamline at the Hamburger Synhrotronstrahlungslabor (HASYLAB) synchrotron radiation facility in Hamburg, Germany. A highly monochromatic pencil X-ray beam in the energy range 9 – 100 keV was used as excitation source. A tunable double Bragg reflection monochromator using a Si[511] and Si[311] set of silicon crystals providing an X-ray resolution of 1 eV at 9 keV rising to 20 eV at 100 keV was used to select the X-ray energies. The beam spot size was set by a pair of precision stepper-driven slits, positioned immediately in front of the sample coupled to the PMT. For all measurements, a slit size of 50 × 50 μm$^2$ was used. The PMT was mounted on an X-Y table capable of positioning with a precision of <1 μm in each direction. Prior to each measurement, the position of the PMT was adjusted to achieve as high count rate as possible. The intensity of the synchrotron beam was reduced in order to avoid pulse pileup. A lead shielding was used to protect the sample from receiving background irradiation which otherwise appeared as a broad background in our pulse height spectra.

To record the synchrotron X-ray pulse height spectra of LaBr$_3$:Ce or LaCl$_3$:Ce, a Hamamatsu R6231-100 PMT connected to a homemade preamplifier, an Ortec 672 spectroscopic amplifier and an Amptek 8000A multichannel analyzer (MCA) were used. The quartz window of the assembly was optically coupled to the window of the PMT with Viscasil 600 000 cSt from General Electric. Corrections were made for channel offsets in the pulse height measurement. The offset was measured by an Ortec 419 precision pulse generator with variable pulse height attenuation settings.

## 3. Results

*3.1. LaBr$_3$:Ce pulse height spectrum*

Figure 1 shows a typical pulse height spectrum recorded with LaBr$_3$:Ce at 45 keV monochromatic X-ray irradiation. The full absorption peak used to determine the photopeak-nPR and the energy resolution is located around channel 710. This peak is a result of the complete deposit of the 45 keV energy of the X-ray photons in the crystal. At channels 167 and 105 lanthanum K$_\alpha$ and K$_\beta$ escape peaks are located. These peaks are the result of X-ray fluorescence escape. X-ray photons of energy between the lanthanum K-electron binding energy $E_{KLa}$=38.925 keV [31] and 100 keV interact with the scintillators almost exclusively by means of the photoelectric effect. After interaction, the electron is ejected from the atom's K-shell, leaving a hole. As the atom returns to its stable lowest energy state, an electron from one of its outer shells jumps into the hole in the K-shell, and in the process giving off a characteristic X-ray photon or Auger electrons. In the case that characteristic X-ray photons escape the bulk of the crystal we observe an escape peak. Since we know precisely the energy of the characteristic X-ray photon the energy deposited in the material is known as well. The procedure has been described by us in detail in [14]. Around channel 530 in figure 1 weak bromine escape peaks can be seen. The amplitude of those peaks is too low so we did not incorporate them in any further analysis.

*3.2. Photon-nPR*

Figure 2 shows the photon-nPR as function of the energy deposited in the bulk of the LaBr$_3$:Ce scintillator while that of the LaCl$_3$:Ce scintillator is shown in figure 3. There are three different types of strongly related photon-nPR curves. The first type is the photopeak-nPR which is derived from a single Gaussian fit of the full absorption peaks in the pulse height spectra recorded with X-ray energies ($E_X$) in the range 9 - 100 keV. We define the nPR of a scintillator at $E_X$ as the number $N_{phe}^{PMT}$/MeV observed at energy $E_X$ divided by the number $N_{phe}^{PMT}$/MeV observed at $E_X$ = 662 keV energy. The nPR is expressed as a percentage value. The second and the third types of photon-nPR curves are the K$_\alpha$-escape-nPR and K$_\beta$-escape-nPR, they are derived from a multi-Gaussian fit of the lanthanum X-ray escape peaks [14]. In order not to blur the data, error bars are only shown for few data points in figures 2 and 3. The typical error in the data for both LaBr$_3$:Ce and LaCl$_3$:Ce is less then 0.05% at 100 keV, rising to 3% at 1.2 keV.

Precision tuning of the X-ray excitation energy at the X-1 beamline at HASYLAB allows us to observe relatively small variations in the photon response near the K, L, and M-shell electron binding energies of the atoms in the compounds. For example for LaBr$_3$:Ce in figure 2 we observe a discontinuity in the photon response curve not only at the lanthanum K-electron binding energy $E_{KLa}$=38.925 keV, but at the bromine K-electron binding energy $E_{KBr}$=13.474 keV [31] as well. The sizes of the jumps in photopeak-nPR are 1.7% and 1.5% the $E_{KLa}$ and $E_{KBr}$ respectively. The total decrease of the photopeak-nPR in the studied range 9 – 100 keV is 15.0%.

The K$_\alpha$-escape-nPR curve of LaBr$_3$:Ce has a dip value of 76.7% at 7.0 keV which is in the energy range above the three lanthanum L-electron shell binding energies of $E_{LLa}$: 5.483 keV, 5.891 keV, and 6.266 keV [31]. The K$_\beta$- escape-nPR curve reaches its minimal value of 68.2% at 2.4 keV which is more then 1 keV above the highest energy lanthanum M-electron shell binding energy of 1.362 keV [31].

The photopeak-nPR curve of LaCl$_3$:Ce as shown in figure 3 has a similar shape as seen for LaBr$_3$:Ce. The curve increases in the energy range from 9 keV to 100 keV by 14.9%. The magnitude of the jump downwards at $E_{KLa}$ is 3.1%. The K$_\alpha$-escape-nPR reaches the lowest value at 6.5 keV and the K$_\beta$-escape-nPR decreases to 54.9% at 2 keV. The photopeak-nPR curves for the two La-halides show similar features as that of the photopeak-nPR curves of LSO:Ce, LuAG:Pr, LPS:Ce and GSO:Ce presented by us in [13].

The attenuation length for X-ray and gamma ray photons in LaBr$_3$ and LaCl$_3$ are also shown in figures 2 and 3. The short attenuation length of low energy X-rays complicates the determination of the photon-nPR of scintillators. These X-rays can be absorbed by air, the beryllium entrance window, the reflector, etc. severely reducing the count rate. More importantly, when X-rays are absorbed within say the first 1 μm, the scintillator light output may be affected by surface effects [6]. By using escape peaks analysis these complications can be avoided.

*3.3. Energy resolution*

The energy resolution $R$ of the X-ray photopeaks for LaBr$_3$:Ce and LaCl$_3$:Ce is plotted on a double-log scale in figure 4 as function of $E_X$. Ideally when only $R_M$ contributes to the energy resolution a straight line with slope -0.5 is expected [9]. For LaBr$_3$:Ce $R$ decreases from 33.4% to 7.2%. A clear step-like increase of almost 1.3% can be seen at energy $E_{KLa}$. A small deviation from a straight line can also be seen at energy around $E_{KBr}$. For LaCl$_3$:Ce in figure 4, $R$ decreases from 41.8% to 8.3%. A step-like increase of 1.1%, analogous to LaBr$_3$:Ce, can be seen around $E_{KLa}$. In the entire range 9 – 100 keV the energy resolution of LaBr$_3$:Ce is better then that of LaCl$_3$:Ce.

Figure 5 shows the energy resolution $R$ versus the number of photoelectrons $N_{phe}^{PMT}$ produced in the Hamamatsu R6231-100 PMT for both scintillators. The solid line represents the theoretical limiting resolution due to the always present Poisson statistics in the number of detected photons, equation (2). The step like increases of resolution at $E_{KLa}$ has actually an "*S-shape*" which can be better seen in the enlarged views on a lin-lin scale of figures 6 and 7. The data point at $E_{KLa}$ is encircled in both of those figures. For both LaBr$_3$:Ce and LaCl$_3$:Ce we observe with increasing $E_X$ that energy resolution starts to increase significantly at 38.8 keV which is approximately 0.1 keV before $E_{KLa}$ is reached. Along with an increase in the resolution, the number of photoelectrons $N_{phe}^{PMT}$ decreases rapidly with the increase in X-ray energy. After further increase of energy by 0.5 keV for LaBr$_3$:Ce and by 1.0 keV for LaCl$_3$:Ce, $N_{phe}^{PMT}$ returns to the value observed at 38.8 keV. We previously observed a similar type of "*S-shape*" behavior for LSO:Ce and other scintillators [13].

*3.4. K-electron-nPR*
Using the K-dip spectroscopy method we derived the K-electron-nPR curves for LaBr$_3$:Ce and LaCl$_3$:Ce which are shown in Figs. 8 and 9. The method is briefly described as follows. An X-ray that photoelectrically interacts with the lanthanum K-shell leads to the creation of a K-shell photoelectron plus several Auger electrons. The response of a scintillator is then equivalent to the sum of two main interaction products: 1) the K-shell photo electron response plus 2) the response from the electrons emitted due to the sequence of processes following relaxation of the hole in the K-shell, the so-called K-cascade response. Our strategy is to employ X-ray energies just above $E_{KLa}$. The K-cascade response is assumed independent from the original X-ray energy. This response is found by tuning the X-ray energy to just above $E_{KLa}$ [12, 16]. By subtracting the K-cascade response from the total X-ray response we are left with the response in photoelectrons from the K-shell photoelectron alone with energy $E_X - E_{KLa}$. The K-electron-nPR curve is then obtained from the number $N_{phe}^{PMT}$/MeV at the energy of the K-photoelectron divided by the number $N_{phe}^{PMT}$/MeV measured at 662 keV. A more detailed description of the K-dip spectroscopy method can be found in [14].

Figure 8 shows the K-electron-nPR of LaBr$_3$:Ce as function of K-photoelectron energy. Across the range 0.07 keV to 61 keV, K-electron-nPR continuously increases from 60% to 96%. The estimated error decreases from ±19% to ±0.05% with increasing energy over the same energy range. Figure 9 shows the K-electron-nPR of LaCl$_3$:Ce as function of K-photoelectron energy. The increase of the K-electron-nPR with increasing K-photoelectron energy for LaCl$_3$:Ce is significantly stronger than for LaBr$_3$:Ce. It rises from 40% at 0.1 keV to 95.3% at 61 keV and the error decreases from ±25% to ±0.05%.

**4. Discussion**

*4.1. Photon-nPR*
The photon-nPR of LaBr$_3$:Ce and LaCl$_3$:Ce shown in figures 2 and 3 are displayed against the deposited amount of energy in the scintillator. This allows us to present the photopeak-nPR, the K$_\alpha$-escape-nPR, and K$_\beta$-escape-nPR curves in one figure. Photopeak-nPR is the standard type of nonproportionality curve that can also be obtained with a set of radioactive sources. In the X-ray energy range from 9 – 100 keV the results match well with the data of other research groups [7, 15, 29, 30] for both scintillators. To extend the nonproportionality curve towards lower energy than 9 keV we performed an analysis of the lanthanum escape peaks to derive the K$_\alpha$-escape-nPR and K$_\beta$-escape-nPR curves. This provides us with the photon-nPR down to energies as low as 1.2 keV for LaBr$_3$:Ce. Figure 2 shows that the K$_\alpha$-escape-nPR curve of LaBr$_3$:Ce does not overlap with the photopeak-nPR curve. The two curves join together only in the small energy range 14.0 -14.6 keV. Below 14.0 keV until 9 keV the photopeak-nPR is almost 4% higher than the K$_\alpha$- escape-nPR. Similarly figure 3 shows that the difference between the photopeak-nPR and the K$_\alpha$-escape-nPR of LaCl$_3$:Ce in the range 9 – 11.6 keV is about 3%. The error of the nPR data in this range is about 0.7% and therefore it can not explain this difference. With $K_\alpha$ or $K_\beta$ X-ray escape, a different set of electrons is generated than without X-ray escape, and although the deposited energy is the same the total light yield will then be different. This is what is seen in figures 2 and 3. When the scintillator response to electrons is available, we expect that all three nPR curves can be reproduced by means of a Monte Carlo simulation of the cascade processes following X-ray interaction in the crystal. It is noted that no significant differences between photopeak-nPR and $K_\alpha$-escape-nPRs were observed for the previously studied inorganic scintillators NaI:Tl [14], LSO:Ce, LuAG:Pr, GSO:Ce and LPS:Ce [13].

Comparison of the results of the photon-nPRs for LaBr$_3$:Ce and LaCl$_3$:Ce show a difference in the magnitude of the drop at $E_{KLa}$. For LaBr$_3$:Ce it is 1.7%, and for LaCl$_3$:Ce it is 3.1%. According to our results for Lu-based materials [13] and calculations by van Loef et al. [6] the magnitude of the drop at the Lutetium K-edge is strongly related to the magnitude of the photon-nPR drop over the entire range. For LSO:Ce, LuAG:Pr and LPS:Ce we observed a proportional dependence between the magnitude of the drop $KL_{drop}$ of scintillator efficiency from below the K-shell to above the L-shell energy of Lu and the magnitude $K_{dip}$ of the drop at the Lu K-edge. It was written as $KL_{drop} = \xi \times K_{dip}$, and empirically we found $\xi \approx 6$. If we use this equation for LaBr$_3$:Ce and LaCl$_3$:Ce we can expect $KL_{drop}$ of 10.2% and 18.6%. The observed $KL_{drop}$ of the photon-nPR is 20.0% for LaBr$_3$:Ce and 22.2% for LaCl$_3$:Ce. Apparently, the relationship is not a rigorous scintillator law.

Figure 4 shows that the energy resolution $R$ of LaBr$_3$:Ce in the entire measurement range 9-100 keV is lower than that of LaCl$_3$:Ce. The 1.3% step-like increase in the energy resolution near $E_{KLa}$ for LaBr$_3$:Ce is higher than the 1.1% increase observed for LaCl$_3$:Ce. This is different from what we observed earlier for LSO:Ce, LuAG:Pr, LPS:Ce and GSO:Ce [13], where the size of the resolution step increases with $K_{dip}$.

*4.2. Energy resolution*

In figure 5 energy resolution is presented as function of the number of created photoelectrons $N_{phe}^{PMT}$. The solid line represents the theoretical contribution due to Poisson statistics, equation (2). Figure5 shows that the energy resolution achieved with LaCl$_3$:Ce as function of the number of detected photons is closer to the statistical limit then that achieved with LaBr$_3$:Ce. However, figure 4 shows that the energy resolution as function of X-ray energy is definitely better for LaBr$_3$:Ce. For LaBr$_3$:Ce the light output is higher and figures 8 and 9 show that it is more proportional. In figure5 statistical contribution $R_M$ goes with $\sqrt{1/N_{phe}^{PMT}}$ whereas the nonproportionality contribution $R_{nPR}$ is not directly related with the $N_{phe}^{PMT}$. Therefore for poor light output scintillators always the statistical contribution dominates. From figure 5 we can conclude that for the high light output crystals the nonproportionality becomes the resolution determining property and this increases the need to estimate the true origin of nonproportionality. LaCl$_3$:Ce is still a crystal where statistics dominates, but for the high output LaBr$_3$:Ce nonproportionality becomes highly important.

The presence of the so-called *S-shape* structures, shown in figures 6 and 7, for LaBr$_3$:Ce and LaCl$_3$:Ce makes these materials not suitable for X-ray spectroscopy in the energy ranges 38.5-39.5 keV and 38.0-40.0 keV respectively. In these ranges there is no unique relationship between $N_{phe}^{PMT}$ and $E_X$. The *S-shape* structures reveal that as the energy of the X-ray photon increases, $N_{phe}^{PMT}$ decreases and the energy resolution deteriorates. The deterioration of the energy resolution starts already at energies approximately 0.1 keV lower then the $E_{KLa}$. This means that some of the processes that cause this deterioration [1, 4, 5, 10, 30] start even before $E_{KLa}$ due to the arising photoabsorption at the lanthanum K-shell electron. At this moment we do not have an explanation for this.

*4.3. Electron-nPR*

Electron-nPRs of LaBr$_3$:Ce and LaCl$_3$:Ce obtained with the K-dip spectroscopy method are shown in figures 8 and 9. Using K-dip spectroscopy we extended the electron response curve down to 70 eV for LaBr$_3$:Ce and down to 100 eV for LaCl$_3$:Ce. We can divide the energy range covered by the K-dip spectroscopy method into three ranges: a) from 61 to 10 keV is a relatively proportional range with slow decrease of scintillator efficiency with decrease of electron energy; b) from 10 to 1 keV there is a fast decrease of scintillator efficiency with decrease of electron energy; and c) below 1 keV there is again like for a) a relatively slow decrease of scintillator efficiency with decrease of electron energy. For LaCl$_3$:Ce shown in Figure 9 this division is somewhat clearer visible than for LaBr$_3$:Ce shown in Figure 8. We already observed similar type of electron nonproportional response curve structure before for LSO:Ce, LuAG:Pr and LPS:Ce [13].

In figures 8 and 9 we have added data for the electron response measured with SLYNCI, the instrument based on the Compton Coincidence Technique (CCT) [18] by Choong et al. [32]. The data gree reasonably well with each other. Like for our data, the SLYNCI data shows that below 10 keV the nPR starts to decrease. However, for both LaBr$_3$:Ce and LaCl$_3$:Ce in the range 10 – 60 keV the SLYNCI- electron-nPR is higher than the K- electron-nPR. It could be caused by different methods of normalization. The SLYNCI- electron-nPR is normalized at 466 keV [32], while our K- electron-nPR is normalized at 662 keV. Furthermore it was assumed by us that the amount of $N_{phe}^{PMT}$ produced by the crystal after absorption of a 662 keV gamma-quantum is equal to the amount produced after absorption of a 662 keV electron [14]. After proper normalization and combining data from SLYNCI with K-dip spectroscopy we aim to obtain a reliable electron-nPR curve in the range 0.07 – 466 keV. By means of Monte Carlo ionization track simulation software we then aim to reproduce the escape-nPR and photopeak-nPR curves of figures 2 and 3.

## 5. Conclusion

The nonproportional scintillation response of LaBr$_3$:Ce$^{3+}$ and of LaCl$_3$:10% Ce$^{3+}$ was measured using highly monochromatic synchrotron irradiation in the energy range 9 – 100 keV. Special attention was paid to the X-ray fluorescence escape peaks as they provide us with additional information about photon response in the range 1.2 - 14.5 keV for LaBr$_3$:Ce and 2.0 - 11.6 keV for LaCl$_3$:Ce. In the X-ray energy range from 9 – 100 keV the results are in a good agreement with the data of other research groups for both scintillators. A rapid variation of the photon response curve is observed near the Lanthanum K- electron binding energy for both scintillators. No relation can be seen between the magnitude of the drop at the Lanthanum K-edge and the magnitude of the photon-nPR drop over the entire range for LaBr$_3$:Ce and LaCl$_3$:Ce.

The presence of the *S-shape* structures in the energy resolution versus $N_{phe}^{PMT}$ curves makes LaBr$_3$:Ce and LaCl$_3$:Ce not suitable for X-ray spectroscopy in the energy ranges 38.5-39.5 keV and 38.0-40.0 keV respectively. In these ranges there is no unique relationship between $N_{phe}^{PMT}$ and $E_X$.

Using K-dip spectroscopy we extended the electron response curve down to 70 eV for LaBr$_3$:Ce and down to 100 eV for LaCl$_3$:Ce. We are not aware of any other experimental method that provides information on electron response to that low energy. Combined data from SLYNCI and K-dip spectroscopy can give us electron-nPR in the entire energy range.

## 6. Acknowledgments

The research leading to these results has received funding from the Netherlands Technology Foundation (STW), Saint Gobain, crystals and detectors division, Nemours, France, and by the European Community's Seventh Framework Programme (FP7/2007-2013) under grant agreement n° 226716. We thank the scientists and technicians of the X-1 beamline at the Hamburger Synhrotronstrahlungslabor (HASY-LAB) synchrotron radiation facilities for their assistance.

## References

[1]. G. Bizarri, W. W. Moses, J. Singh, A. N. Vasil'ev, *J. Appl. Phys.* **105** (2009) 044507.
[2]. P. Dorenbos, J. T. M. De Haas, and C. W. E. Van Eijk, *IEEE Trans. Nucl. Sci.* **42** (1995) 2190.
[3]. J.E. Jaffe, D.V. Jordan, A.J. Peurrung, *Nucl. Instr. Meth. A.* **570** (2007) 72.
[4]. W.W. Moses S. A. Payne, W.-S. Choong, G. Hull, and B. W. Reutter, *IEEE Trans. Nucl. Sci.* **55** (2008) 1049.
[5]. P. A. Rodnyi, P. Dorenbos, C. W. E. van Eijk, *Phys. Stat. Sol. (b)* **187** (1995) 15.
[6]. E. V. D. van Loef, W. Mengesha, J. D. Valentine, P. Dorenbos, C. W. E. van Eijk, *IEEE Trans. Nucl. Sci.* **50** (2003) 155.
[7]. M. Moszynski, Ł.S widerski, T. Szczesniak, A. Nassalski, A. Syntfeld-Kazuch, W. Czarnacki, G. Pausch, J. Stein, P. Lavoute, F. Lherbert, F. Kniest, *IEEE Trans. Nucl. Sci.* **55** (2008) 1774.
[8]. P. Dorenbos, *Nucl. Instr. Meth. A.* **486** (2002) 208.
[9]. P. Dorenbos, *IEEE Trans. Nucl. Sci.* **57** (2010) 1162.
[10]. A.N.Belsky, R.A.Glukhov, I.A.Kamenskikh, P.Martin, V.V.Mikhailin, I.H.Munro, C. Pedrini, D.A.Shaw, I.N.Shpinkov, A.N.Vasil'ev, *J. Electr. Spectr. Rel. Phen.* **79** (1996) 147.
[11]. P. A. Rodnyi, *Rad. Measur.* **29** (1998) 235.
[12]. A. J. L. Collinson, R. Hill, *Proc. Phys. Soc.* **81** (1963) 883.
[13]. I.V. Khodyuk, J.T.M. de Haas, P. Dorenbos, *IEEE Trans. Nucl. Sci.* **57** (2010) 1175.
[14]. I.V. Khodyuk, P.A. Rodnyi, *P. Dorenbos, J. Appl. Phys.* **107** (2010) 113513.
[15]. A. Owens, A.J.J. Bos, S. Brandenburg, P. Dorenbos, W. Drozdowski, R.W. Ostendorf, F. Quarati, A. Webb, E. Welter, *Nucl. Instr. Meth. A.* **574** (2007) 158.
[16]. L.R. Wayne, W.A. Heindl, P.L. Hink, R.E. Rothschild, *Nucl. Instr. and Meth. A.* **411** (1998) 351.
[17]. M. Moszynski, M. Balcerzyk, W. Czarnacki, M. Kapusta, W. Klamra, A. Syntfeld, M. Szawlowski, *IEEE Trans. Nucl. Sci.* **51** (2004) 1074.
[18]. J. D. Valentine, B. D. Rooney, *Nucl. Instr. Meth. A.* **353** (1994) 37.
[19]. J. Andriessen, O.T. Antonyak, P. Dorenbos, P.A. Rodnyi, G.B. Stryganyuk ,C.W.E. van Eijk, A.S. Voloshinovskii, *Opt. Comm.* **178** (2000) 355.
[20]. E. V. D. van Loef, P. Dorenbos, C. W. E. van Eijk, K. Krämer, and H. U.Güdel, *Appl. Phys. Lett.* **77** (2000) 1467.
[21]. E. V. D. van Loef, P. Dorenbos, C. W. E. van Eijk, K. Krämer, and H. U.Güdel, *Appl. Phys. Lett.* **79** (2001) 1573.
[22]. A. Iltis, M.R. Mayhugh, P. Menge, C.M. Rozsa, O. Selles, V. Solovyev, *Nucl. Instr. Meth. A.* **563** (2006) 359.
[23]. W. W. Moses, K. S. Shah, *Nucl. Instr. Meth. A.* **537** (2005) 317.
[24]. N.J. Cherepy, S.A. Payne, S.J. Asztalos, G. Hull, J.D. Kuntz, T. Niedermayr, S. Pimputkar, J.J. Roberts, R.D. Sanner, T.M. Tillotson, E. van Loef, C.M. Wilson, K.S. Shah, U.N. Roy, R. Hawrami, A. Burger, L.A. Boatner, W.-S. Choong, W.W. Moses, *IEEE Trans. Nucl. Sci.* **56** (2009) 873.
[25]. D. R. Schaart, S. Seifert, R. Vinke, H. T van Dam, P. Dendooven, H. Lohner, F. J. Beekman, *Phys. Med. Biol.* **55** (2010) N179.
[26]. L. Swiderski, M. Moszyn´ski, A. Nassalski, A. Syntfeld-Kazuch, T. Szczes´niak, K. Kamada, K. Tsutsumi, Y. Usuki, T. Yanagida, A. Yoshikawa, W. Chewpraditkul, M. Pomorski, *IEEE Trans. Nucl. Sci.* **56** (2009) 2499.
[27]. J. T. M. de Haas, P. Dorenbos, *IEEE Trans. Nucl. Sci.* **55** (2008) 1086.
[28]. S. Kraft, E. Maddox, E.-J. Buis, A. Owens, F. G. A. Quarati, P. Dorenbos, W. Drozdowski, A. J. J. Bos, J. T. M. de Haas, H. Brouwer, C. Dathy, V. Ouspenski, S. Brandenburg, R. Ostendorf, *IEEE Trans. Nucl. Sci.* **54** (2007) 873.
[29]. C. D'Ambrosio, F. de Notaristefani, G. Hull, V. O. Cencelli, R. Pani, *Nucl. Instr. Meth. A.* **556** (2006) 187.
[30]. S. A. Payne, N. J. Cherepy, G. Hull, J. D. Valentine, W. W. Moses, W.-S. Choong, *IEEE Trans. Nucl. Sci.* **56** (2009) 2506.
[31]. X-Ray Data Booklet http://xdb.lbl.gov/
[32]. W.-S. Choong, K. M. Vetter, W. W. Moses, G. Hull, S. A. Payne, N. J. Cherepy, J. D. Valentine, *IEEE Trans. Nucl. Sci.* **55** (2008) 1753.

Figure 1 Pulse height spectrum measured with LaBr$_3$:Ce at 45 keV monochromatic X-ray irradiation.
Figure 2 Photon nonproportional response of LaBr$_3$:Ce as a function of deposited energy. Black solid circles, photopeak-nPR; blue open squares, K$_\alpha$ escape-nPR; red open circles, K$_\beta$ escape-nPR .The solid curve shows the calculated X-ray attenuation length for LaBr$_3$.
Figure 3 Photon nonproportional response of LaCl$_3$:Ce as a function of deposited energy. Black solid circles, photopeak-nPR; blue open squares, K$_\alpha$ escape-nPR; red open circles, K$_\beta$ escape-nPR .The solid curve shows the calculated X-ray attenuation length for LaCl$_3$.
Figure 4 Energy resolution of the X-ray photopeak as a function of X-ray energy. LaBr$_3$:Ce – black open squares, LaCl$_3$:Ce – red solid circles.
Figure 5 Energy resolution as function of the number of photoelectrons $N_{phe}^{PMT}$. LaBr$_3$:Ce – black open squares, LaCl$_3$:Ce – red solid circles. Solid line – contribution predicted from Poisson statistics.
Figure 6 S type structure near the Lanthanum K-electron binding energy $E_{KLa}$=38.925 keV for LaBr$_3$:Ce.
Figure 7 S type structure near the Lanthanum K-electron binding energy $E_{KLa}$=38.925 keV for LaCl$_3$:Ce.
Figure 8 Black solid circles, electron nonproportional response of LaBr$_3$:Ce as a function of K-photoelectron energy obtained from K-dip spectroscopy. Red open squares, electron-nPR obtained with SLYNCI from [7].
Figure 9 Black solid circles, electron nonproportional response of LaCl$_3$:Ce as a function of K-photoelectron energy obtained from K-dip spectroscopy. Red open squares, electron-nPR obtained with SLYNCI from [7].

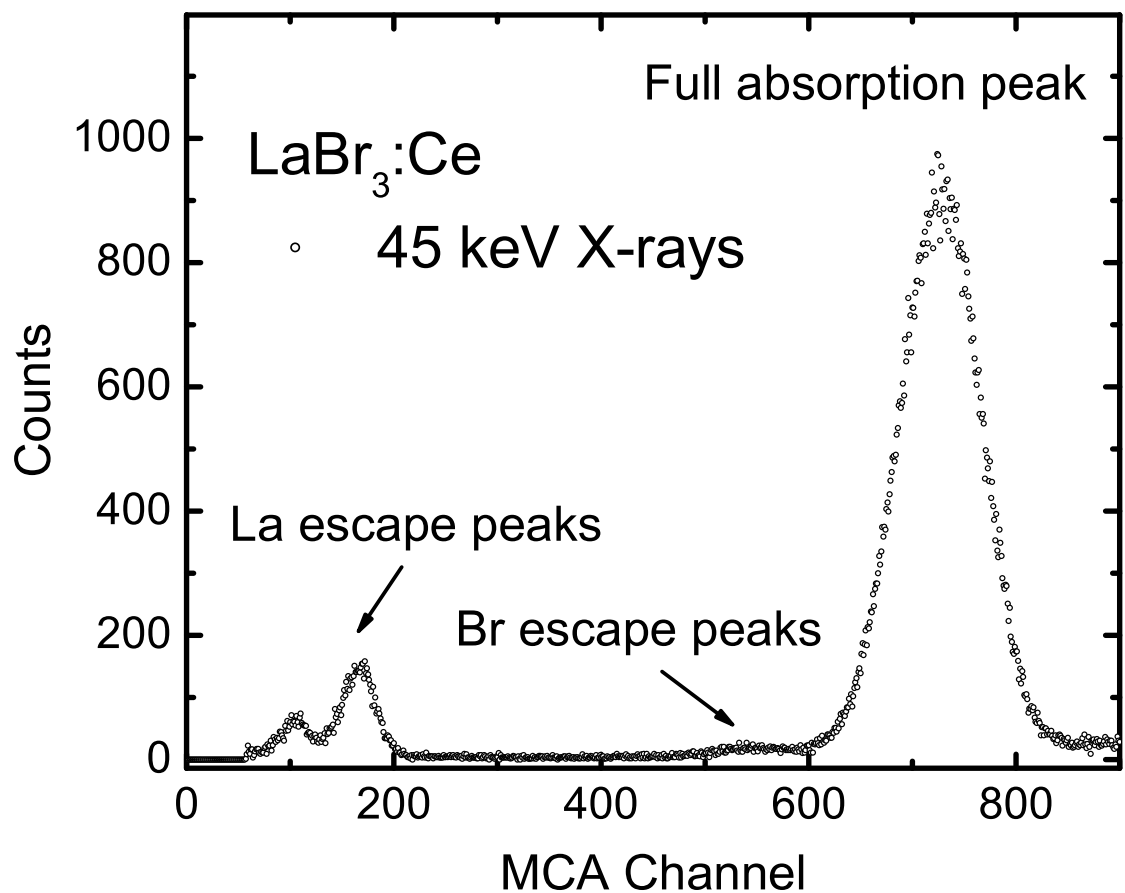

Figure 1 (Figure1.eps)

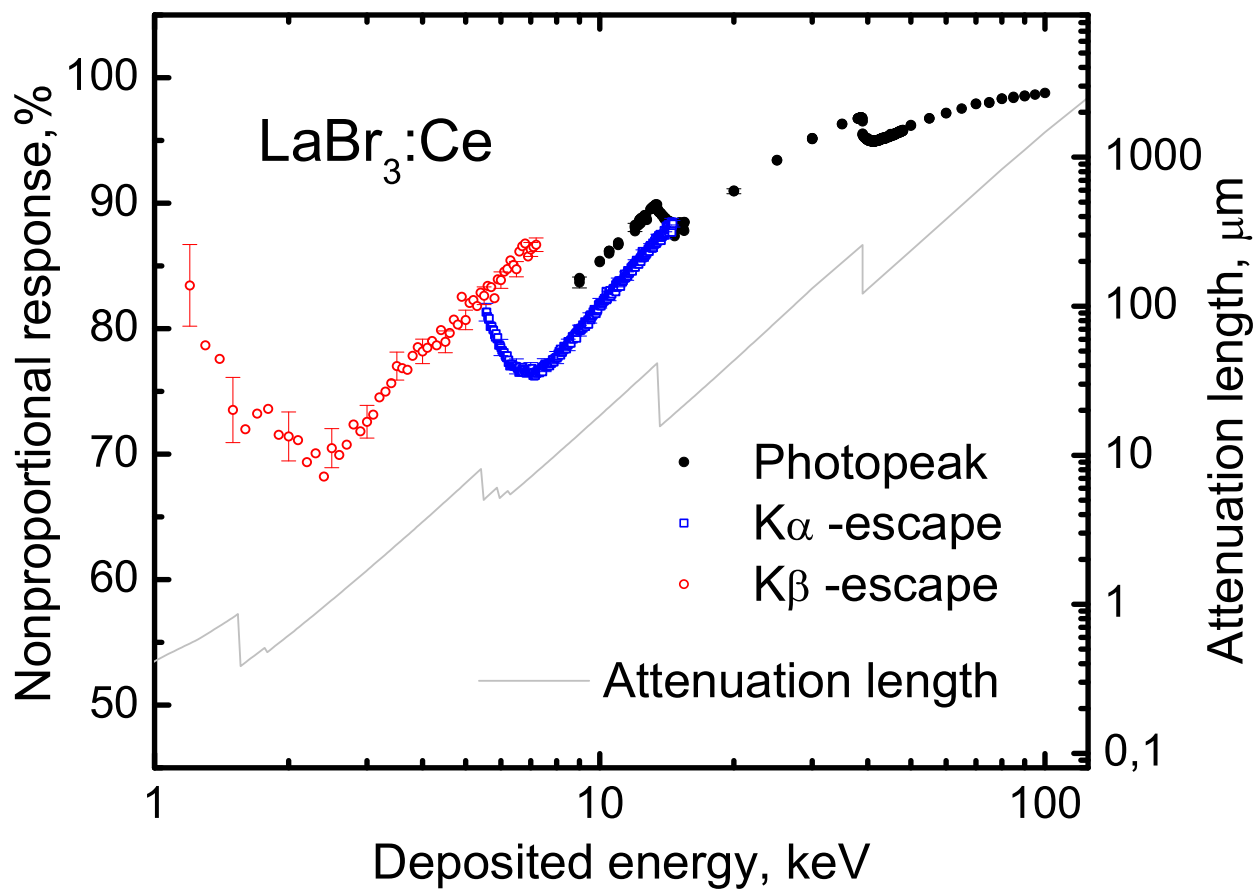

Figure 2 (Figure2.eps)

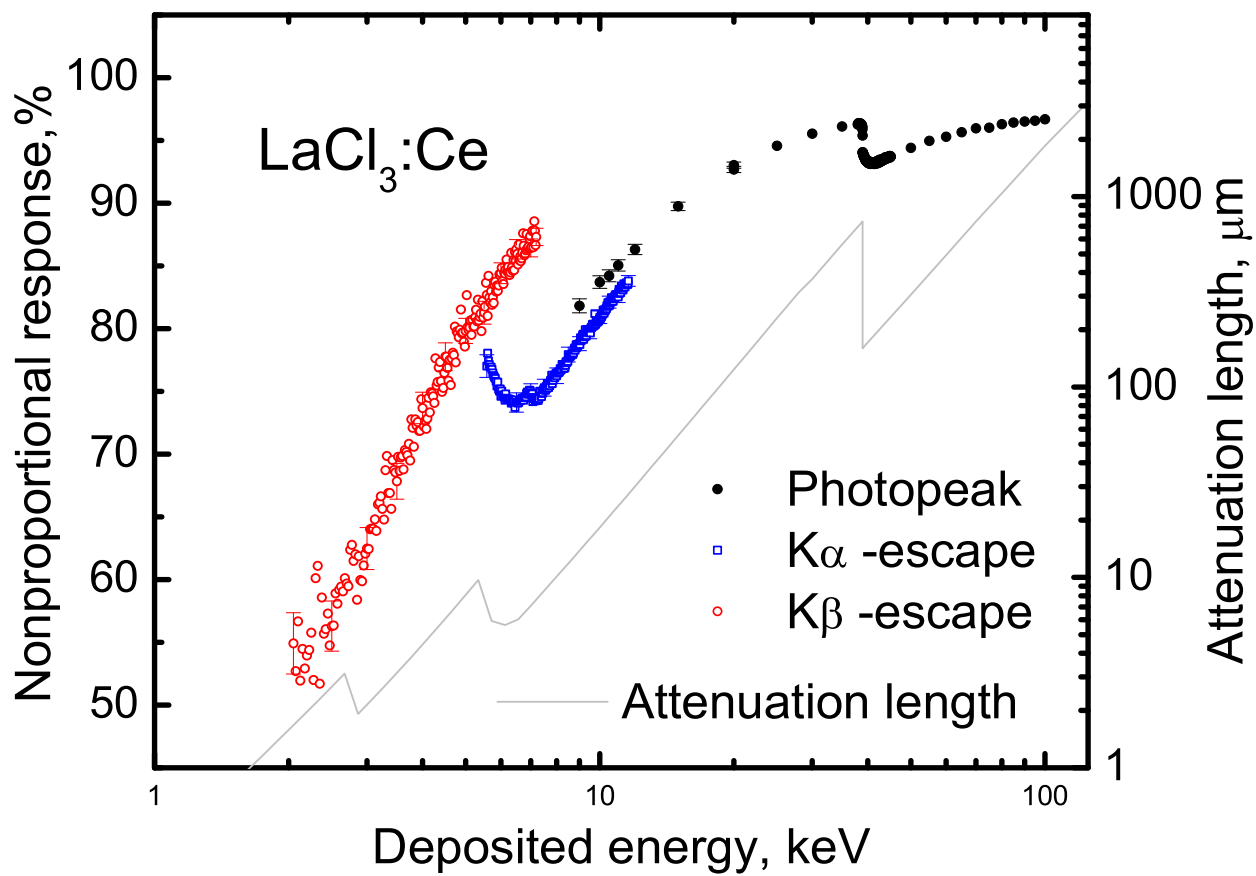

Figure 3 (Figure3.eps)

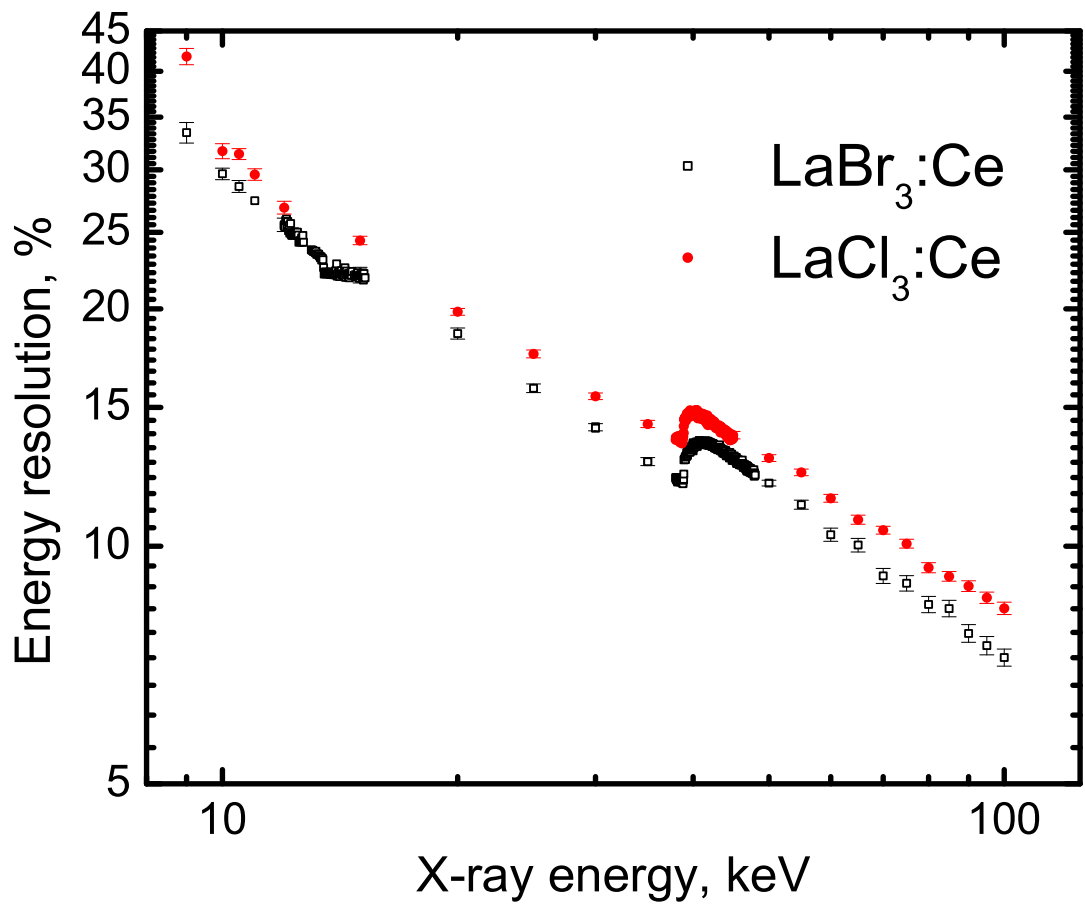

Figure 4 (Figure4.eps)

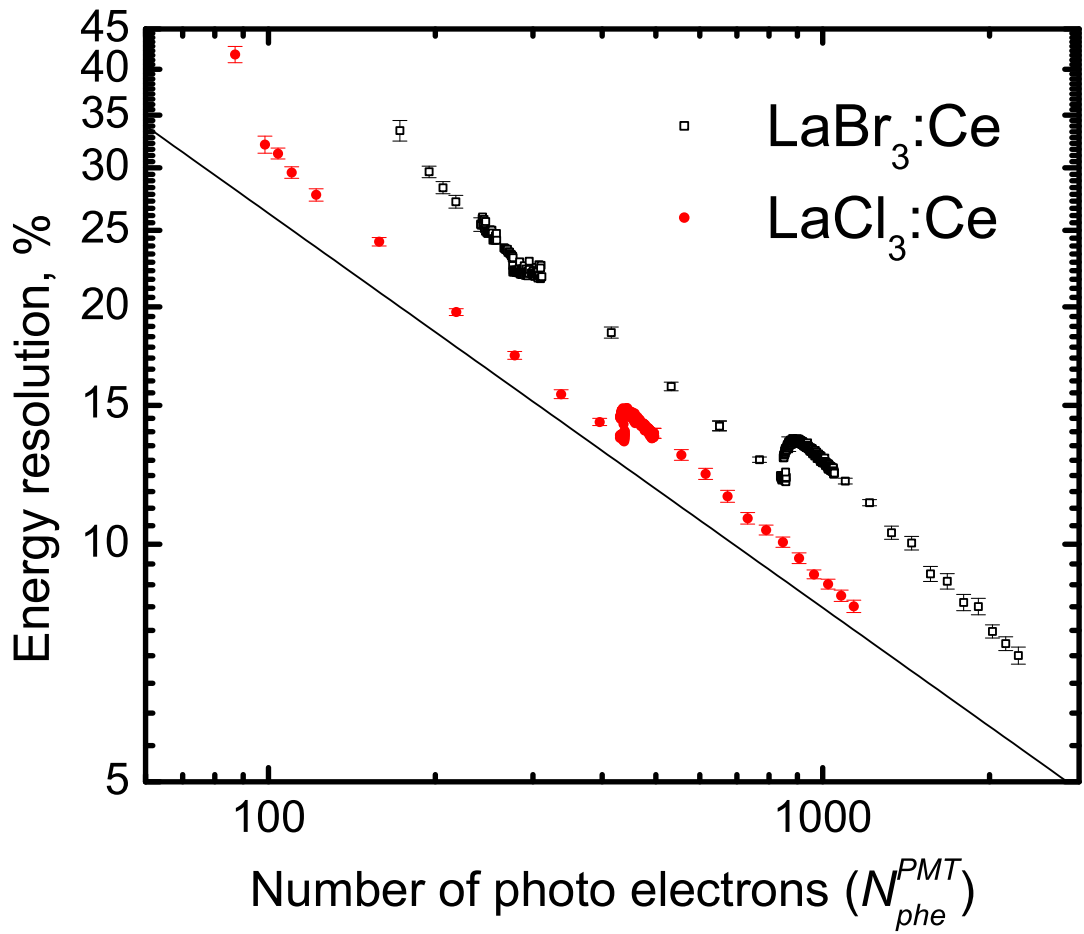

Figure 5 (Figure5.eps)

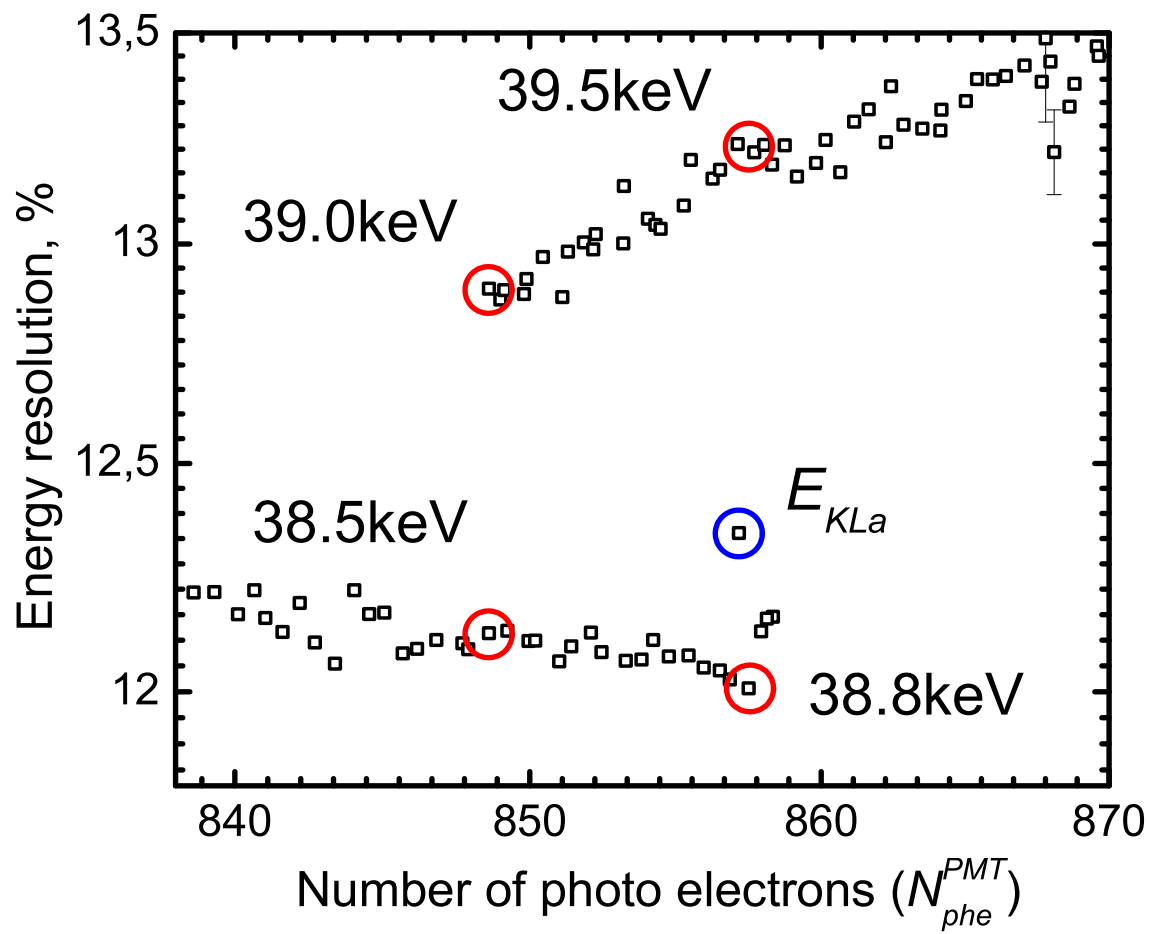

Figure 6 (Figure6.eps)

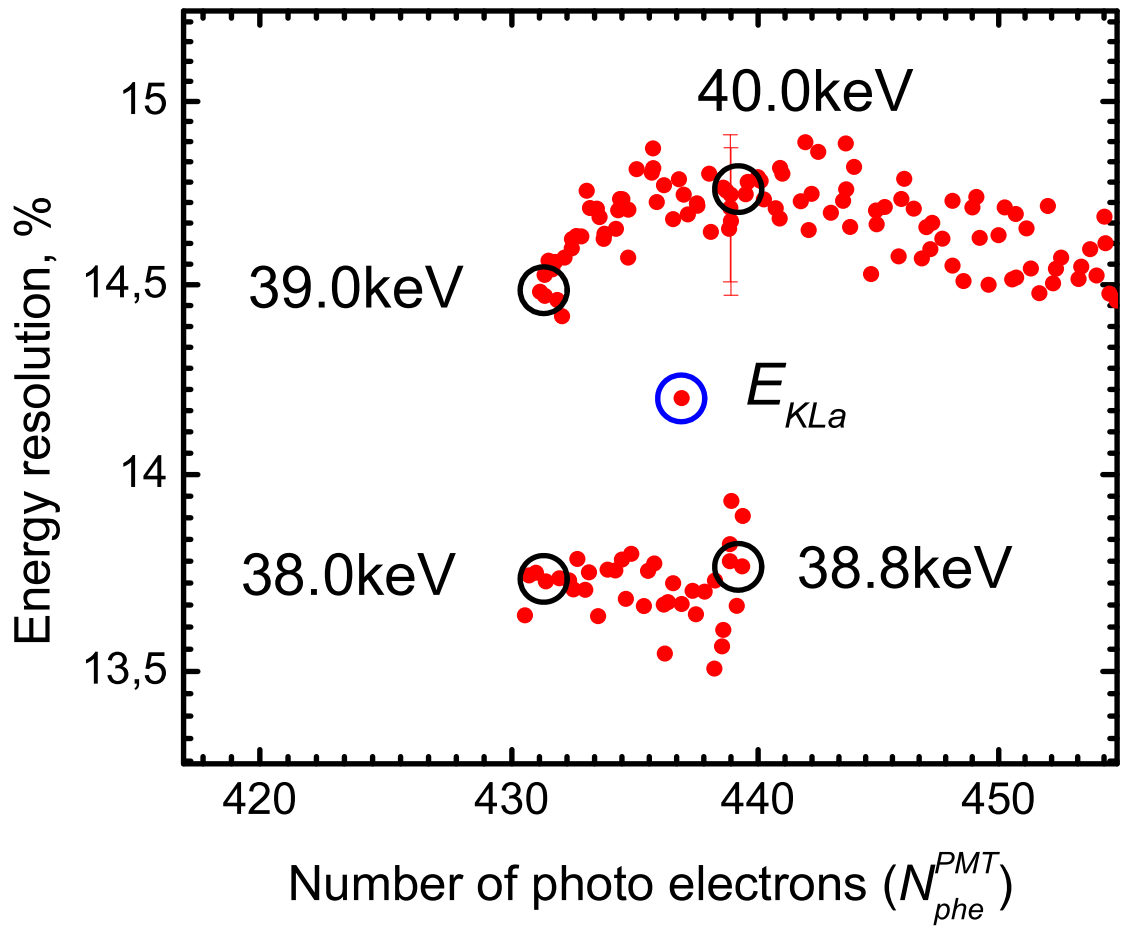

Figure 7 (Figure7.eps)

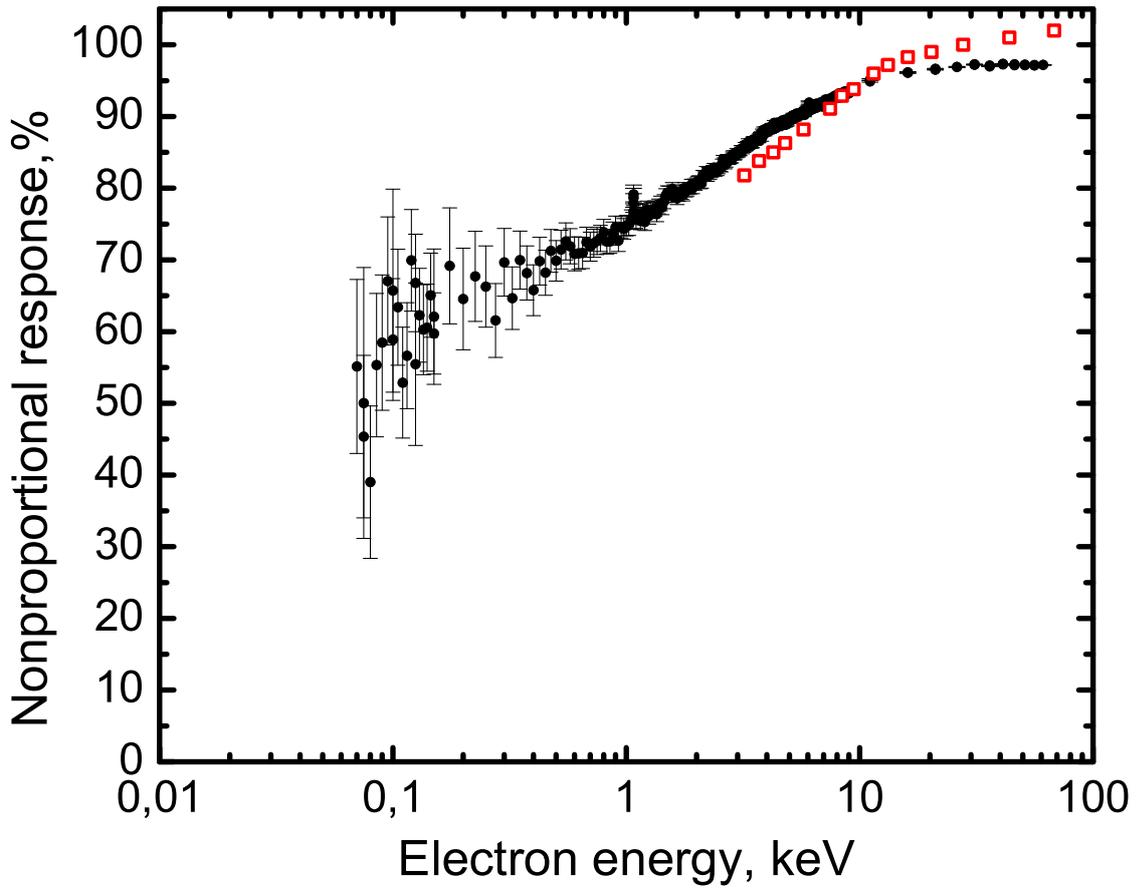

Figure 8 (Figure8.eps)

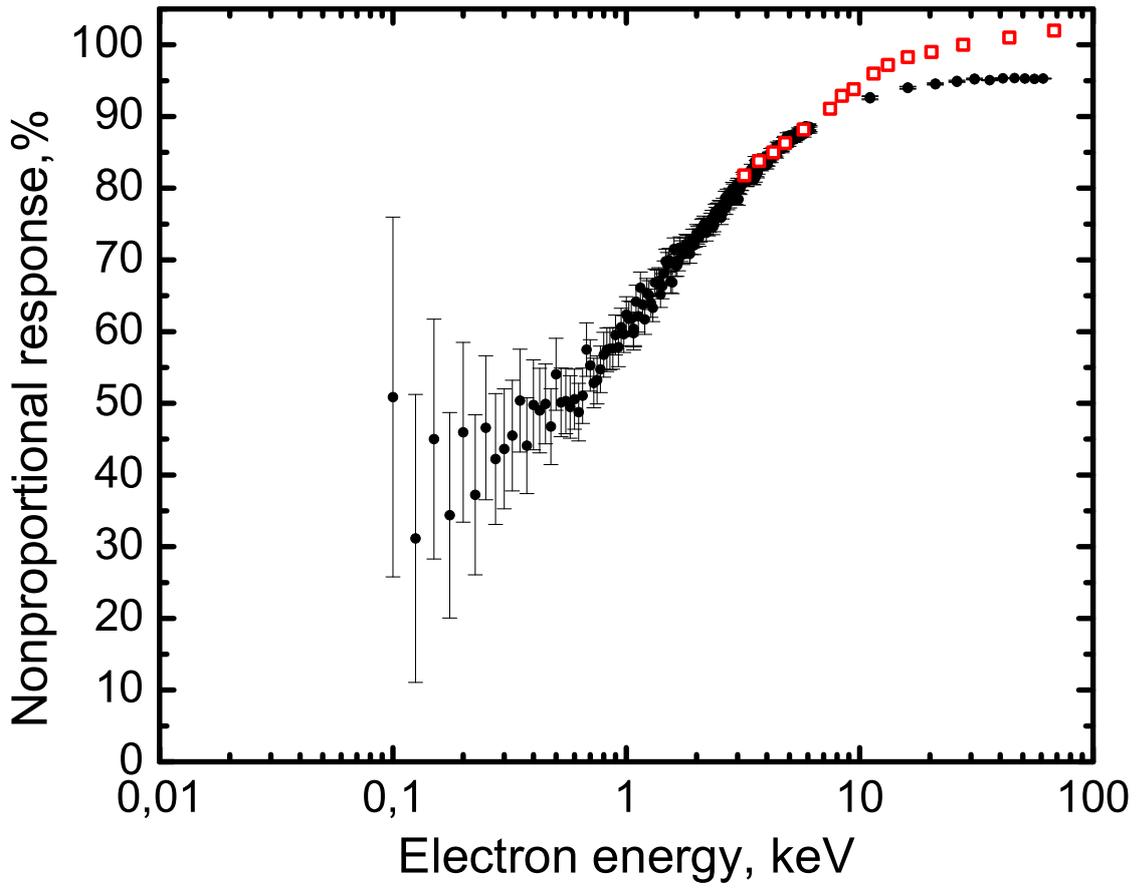

Figure 9 (Figure9.eps)